\documentclass[aps,prc,twocolumn,superscriptaddress,breaklinks,showpacs]{revtex4-1}

\usepackage{bm}
\usepackage[dvips]{graphicx}
\usepackage{delarray}
\usepackage{array}
\usepackage{amsbsy}
\usepackage{color} 
\usepackage{pifont}
\usepackage{float} 
\usepackage{indentfirst} 
\usepackage{hyperref} 
\usepackage{amssymb} 
\usepackage{setspace} 
\usepackage{ulem} 
\newcommand{\capn}{$^{11}$C$(\alpha,p)^{14}$N}
\newcommand{\npac}{$^{14}$N$(p,\alpha)^{11}$C}
\newcommand{\alphap}{$(\alpha,p)$}
\newcommand{\apo}{$(\alpha,p_0)$}

\newcommand{\api}{$(\alpha,p_1)$}
\newcommand{\apii}{$(\alpha,p_2)$}
\newcommand{\capno}{$^{11}$C$(\alpha,p_0)^{14}$N}
\newcommand{\capni}{$^{11}$C$(\alpha,p_1)^{14}$N$^\ast$}
\newcommand{\capnii}{$^{11}$C$(\alpha,p_2)^{14}$N$^\ast$}
\newcommand{\ca}{$^{11}$C~+~$\alpha$}

\newcommand{\cm}{_{\rm c.m.}}
\newcommand{\reac}[6]{$^{#1}$#2$(#3,#4)^{#5}$#6}
\begin{document}


\title{First direct measurement of the 
 \capn\ stellar reaction by an extended thick-target method}


\author{S.~Hayakawa}
\email{hayakawa@cns.s.u-tokyo.ac.jp}
\affiliation{Center for Nuclear Study, 
 the University of Tokyo, RIKEN campus, 2-1 Hirosawa,Wako, Saitama 351-0198, Japan}
\author{S.~Kubono}
\affiliation{Center for Nuclear Study, 
 the University of Tokyo, RIKEN campus, 2-1 Hirosawa,Wako, Saitama 351-0198, Japan}
\affiliation{RIKEN Nishina Center,
 2-1 Hirosawa, Wako, Saitama 351-0198, Japan}
\affiliation{Institute of Modern Physics, Chinese Academy of Science, 
Nanchang Road 509, Lanzhou 730000, People's Republic of China}
\author{D.~Kahl}
\altaffiliation[Present address: ]{
School of Physics and Astronomy, the University of Edinburgh, James Clerk Maxwell Building, Peter Guthrie Tait Road, Edinburgh, EH9 3FD, UK}
\affiliation{Center for Nuclear Study, 
 the University of Tokyo, RIKEN campus, 2-1 Hirosawa,Wako, Saitama 351-0198, Japan}
\author{H.~Yamaguchi}
\affiliation{Center for Nuclear Study, 
 the University of Tokyo, RIKEN campus, 2-1 Hirosawa,Wako, Saitama 351-0198, Japan}
\author{D.~N.~Binh}
\affiliation{Institute of Physics, 
 Vietnamese Academy for Science and Technology, 
 10 Daotan, Thule, Badinh, Hanoi, Vietnam}
\author{T.~Hashimoto}
\affiliation{Rare Isotope Science Project, 
 Institute for Basic Science, 
 Yuseong-gu Daejeon 305-811, Republic of Korea}
\author{Y.~Wakabayashi}
\affiliation{RIKEN Nishina Center,
 2-1 Hirosawa, Wako, Saitama 351-0198, Japan}
\author{J.~J.~He}
\affiliation{Institute of Modern Physics, Chinese Academy of Science, 
Nanchang Road 509, Lanzhou 730000, People's Republic of China}
\author{N.~Iwasa}
\affiliation{Department of Physics, Tohoku University, 
 Aoba, Sendai, Miyagi 980-8578, Japan}
\author{S.~Kato}
\affiliation{Department of Physics, Yamagata University, 
 1-4-12 Kojirakawa-machi, Yamagata 990-8560, Japan}
\author{T.~Komatsubara}
\affiliation{RIKEN Nishina Center,
 2-1 Hirosawa, Wako, Saitama 351-0198, Japan}
\author{Y.~K.~Kwon}
\affiliation{Rare Isotope Science Project, 
 Institute for Basic Science, 
 Yuseong-gu Daejeon 305-811, Republic of Korea}
\author{T.~Teranishi}
\affiliation{Department of Physics, Kyushu University, 
744 Motooka, Nishi-ku, Fukuoka 819-0395, 
Japan}


\date{\today}

\begin{abstract}
 The \capn\ reaction is 
 an important $\alpha$-induced reaction
 competing with $\beta$-limited hydrogen-burning processes
 in high-temperature explosive stars.
 We directly measured its reaction cross sections 
 both for the ground-state transition \apo\
 and the excited-state transitions \api\ and \apii\
 at relevant stellar energies 
 1.3--4.5~MeV
 by an extended thick-target method featuring time of flight
 for the first time. 
 %
 %
 We revised the reaction rate by numerical integration
 including the \api\ and \apii\ contributions and also
 low-lying resonances of \apo\ using both the present and the
 previous experimental data which were totally neglected in the previous
 compilation works. 
 %
 The present total reaction rate lies 
 between the previous \apo\ rate 
 and the total rate of the Hauser-Feshbach statistical model
 calculation, which is consistent with the relevant 
 explosive hydrogen-burning scenarios such as the $\nu p$-process.
\end{abstract}

\pacs{29.30.Ep,26.30.-k,24.30.-v,25.55.-e}

\maketitle


%
\section{Introduction}
Hydrogen burning at high temperatures 
of the order of $T_9 \sim 0.1$--1 ($T_9$ represents 1~GK)
is of great importance in terms of 
production of heavy elements beyond iron and energy generation
in various high-temperature and explosive astrophysical sites,
such as supermassive metal-poor stars,
type-I 
x-ray bursts, and core-collapse supernovae.
The nucleosynthesis paths and the 
timescale of the 
$rp$-process in such stellar environments depend on the  
simultaneously occurring  
$\alpha p$-process \cite{wal-woo1981}.
This process breaks out from the hot CNO cycle 
and bypasses slower $\beta$-decays
by sequences of \alphap\ reactions and proton captures 
up to mass number $A \sim 40$,
drastically depending on temperature.
%
In type-I x-ray bursts, which are
the most frequent nuclear explosion in the universe \cite{Fisker2008}, 
the $\alpha p$-process may determine
the time evolution of energy release which varies 
by 2--3 orders of magnitude in about 1 second,
and thus several key \alphap\ reactions,
{\it e.g.} \reac{14}{O}{\alpha}{p}{17}{F} and 
\reac{18}{Ne}{\alpha}{p}{21}{Na}, have been investigated 
\cite{Mohr2013}.
Some studies for metal-poor stars \cite{wiescher1989hot}
and the $\nu p$-process \cite{WanajoJankaKubono2011}
suggest that several $\alpha$-induced reaction sequences,
such as \reac{7}{Be}{\alpha}{\gamma}{}{}\capn\
and \reac{7}{Be}{\alpha}{p}{10}{B}\reac{}{}{\alpha}{p}{13}{C},
also may appear significant 
in $T_9 = 1.5$--3.
%
These sequences are shown to
bridge the mass gap at $A = 8$ 
comparably to the triple-$\alpha$ process,
which may even affect the abundances of $p$-nuclei
around $A = 90$ \cite{WanajoJankaKubono2011}.
%

The current problem 
in nucleosynthesis is that most of the relevant
\alphap\ reaction rates have been estimated only through 
studies on resonances, time-reversal reactions,
or other indirect methods, with only a
few direct measurements 
\cite{PhysRevLett.84.1651, PhysRevC.59.3402, Groombridge2002, NSR2004NO18}.
%
These direct measurements provided limited results
due to experimental difficulties, such as detection efficiency,
production of radioactive ion beams,
determination of reaction points in the $^4$He gas target,
identification of the final excited states, etc.
Recently, some research projects for direct measurement of 
\alphap\ reactions of astrophysical interest are ongoing
using an ambitious detection 
system such as HELIOS (Helical Orbit Spectrometer
\cite{Wuosmaa2007, Lighthall2010}),
which has been used for $(d,p)$ reaction measurements, 
yet under commissioning
for \alphap\ reaction measurements.
%
Besides these works,
we have successfully 
performed a direct measurement of the \capn\ reaction
for the first time
by a simple experimental configuration,
leading one of the most comprehensive results
among the above crucial nucleosynthesis paths.
%
%
\\ \indent
The \capn\ reaction rates currently available 
are reported in compilations by 
Caughlan and Fowler \cite{caughlan1988thermonuclear} (hereafter ``CF88''),
and NACRE collaboration \cite{nacre99}.
Both of them are based on studies of the time-reversal reaction \npac\ 
by the activation method \cite{Shrivastava1968, PhysRevC.9.2134,
CasellaV.R.ChristmanD.R.1978, Epherre1971, 
PhysRevC.13.524, G.T.BidaT.J.Ruth1980},
which thus provide information only on the transition
to the ground state of $^{14}$N 
(hereafter ``\apo''), but 
not to the excited states (hereafter ``\api'', ``\apii'', etc.).
%
Indeed, in nuclear astrophysics, transitions to excited states are
rarely investigated experimentally in spite of their importance.
In the CF88 compilation, the astrophysical $S$-factor
of this reaction as well as that of 
the time-reversal reaction
were adopted from the lowest-energy study
by Ingalls {\it et~al.} ~\cite{PhysRevC.13.524} 
which approximated the cross section over several resonances
by a smooth function.
The NACRE compilation derived the 
$S$-factor
of the time-reversal reaction
from 6 data sets \cite{Shrivastava1968, PhysRevC.9.2134,
CasellaV.R.ChristmanD.R.1978, Epherre1971, 
PhysRevC.13.524, G.T.BidaT.J.Ruth1980}
with improvement at higher energies,
but it obviously underestimated the lowest energy part
which is important for the reaction rate near $T_9$ temperatures.
An alternative compilation work on the \npac\ reaction 
by Tak\'acs {\it et~al.} \cite{Takacs2003169}
has applied the Pad\'e approximation to the excitation functions 
of 13 data sets up to 
$E\cm \sim 25$~MeV
in the \ca\ center-of-mass system
reproducing resonances more properly,
but the resonances below $E\cm = 1.5$~MeV  were again
just smoothed out.
%
A direct measurement is able to address the above omissions 
in the reaction rate evaluations, and also
enables us to observe such excited-state transitions
and also has an advantage 
to validate the cross sections measured by the activation method.
%
\section{Experimental method and data analysis}
\begin{figure}[t!] 
 \centering
 \resizebox{8.3cm}{!}{
 \includegraphics{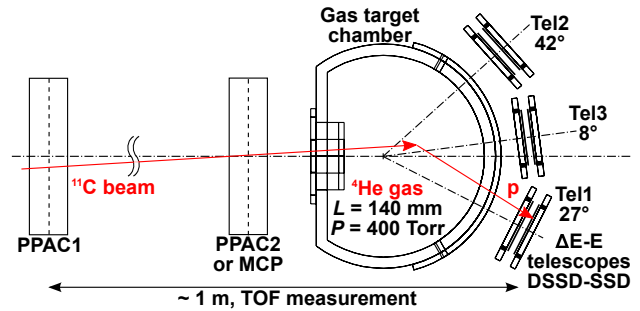}}
 \caption{\label{fig:setup}
 (Color online) Plane view of the experimental setup.
 %
 }
\end{figure}

The experiment was performed in inverse kinematics
using the unstable nuclide $^{11}$C as the projectile
covering a center-of-mass energy range 0--4.5~MeV
based on the thick target method 
\cite{NSR1990AR24}.
%
The $^{11}$C beams were produced by the in-flight technique
with CRIB 
(Center for Nuclear Study Radioactive 
Ion Beam separator \cite{yanagisawa2005low}).
A $^{11}$B$^{3+}$ primary beam at 4.6~MeV/nucleon  with a typical 
intensity of 1~p$\mu$A was provided by the RIKEN AVF cyclotron, and 
bombarded the cryogenic hydrogen 
gas target \cite{yamaguchi2008development}
with a typical thickness of 1.7~mg/cm$^2$, 
confined by 2.5-$\mu$m-thick Havar foils at beam entrance and exit.
The secondary $^{11}$C ions 
produced via the 
$^1$H$(^{11}$B,$^{11}$C$)n$
reaction were purified by the double achromatic system and
the following Wien filter of CRIB. 
\\ \indent
The experimental setup at the final focal plane,
illustrated in Fig.~\ref{fig:setup},
consisted of two beam-line monitors,
a $^4$He gas target, and 
three sets of 
$\Delta$E-E position-sensitive 
silicon detector telescopes.
%
%
%
\begin{figure}[t!]
 \centering
 \resizebox{8.5cm}{!}{
 \includegraphics{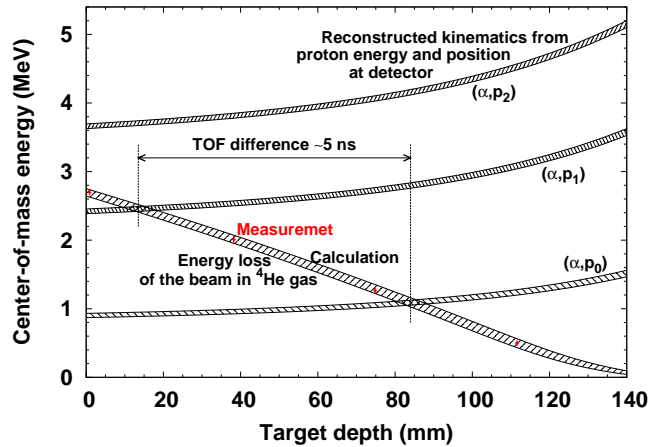}}
 \caption{\label{fig:ebeam_ep}
 (Color online) 
 Example of 
 the center-of-mass energies as functions of 
 the target depth.
 The decreasing function is determined from the 
 energy loss calculation
 with the incident beam energy of $10.12\pm0.46$~MeV,
 which is consistent with
 the actual measurement shown as red crosses. 
 The increasing functions are reconstructed by kinematic calculations 
 and energy loss corrections of a detected proton
 for \apo, \api\ and \apii\ with a given kinetic energy of 5~MeV and
 a scattering angle of 40$^{\circ}$.
 The vertices of these functions
 represent the possible reaction energy and position.
 %
 }
\end{figure}
%
A Parallel Plate Avalanche Counter (PPAC) \cite{kumagai2001delay} 
was used for the first beam-line monitor,
and another PPAC and a MicroChannel Plate detector (MCP)
\cite{hay2008}
were switchable for the second monitor; 
the employed experimental setup allowed us
to alternate between two different on-target energies 
by the 
different thicknesses
of the second beam-line monitors,
thus scanning a wide excitation-energy
range in two measurements
without altering the CRIB optical condition.
%
This setup
was also helpful to confirm self-consistency of our measurement
by checking the data in the overlapping energy region
of the lower- and the higher-energy run,
as mentioned later.
%
These two monitors 
tracked each beam particle event-by-event,
and the first monitor also acted as the time reference
of the measurement.
The data acquisition was triggered by 
the sum of the downscaled signal of the first PPAC
and the coincident signal of the first PPAC with the silicon detectors,
which thus canceled out the PPAC efficiency for the absolute cross section
determination.
The two resulting beam energies 
were directly measured with a silicon detector
at the target position, to be 
10.12$\pm$0.46~MeV and 16.86$\pm$0.36~MeV 
in full width at half maximum, 
covering center-of-mass energy ranges 
0--2.7~MeV and 2.3--4.5~MeV, 
respectively.
%
The secondary-beam ions were identified event by event
from the time of flight (TOF) information between the two beam-line monitors
and the TOF with respect to the cyclotron radio-frequency signal,
thus the beam contaminants were easily 
distinguished in the analysis.
The $^{11}$C$^{6+}$ purity was 
determined to be higher than 97\%, 
and the rest was mostly $^{11}$B$^{5+}$ 
and trace amounts of $^{11}$C$^{5+}$.
%
The $^{11}$C beam spot was well focused on the target window 
with a diameter of 30~mm so that 
78\% of the $^{11}$C beam particles eventually entered the target chamber,
and the on-target average intensities 
of the lower- and the higher-energy $^{11}$C beams were 
3.1$\times$10$^{5}$~pps and 1.0$\times$10$^{5}$~pps, 
respectively.
\\ \indent
The $^4$He gas target was sealed with 2.5-$\mu$m-thick Havar foil
at the entrance window, 
and 25-$\mu$m-thick Mylar foil at the exit window
which is in a cylindrical shape and has a sufficiently large area
for the full detection acceptance.
%
The energy straggling of the ejected proton through this exit window is 
estimated at only 20~keV at most, which is negligible.
%
%
\begin{figure}[t!] 
 \centering
 \resizebox{7.7cm}{!}{
 \includegraphics{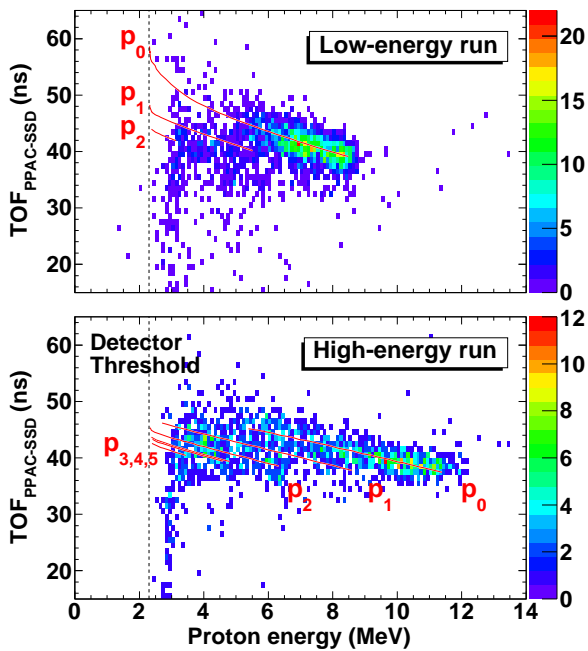}}
 \caption{\label{fig:t-e_high-e_bg}
 (Color online) TOF from the first PPAC to the SSD versus 
 energy of protons detected in
 Tel1 of 
 the lower-energy run (upper)
 and the higher-energy run (lower) 
 with calculated lines for allowed 
 excited-state 
 transitions.
 }
\end{figure}
%
\\ \indent
The silicon telescopes were mounted at three different laboratory angles,
27$^{\circ}$, 42$^{\circ}$ and 8$^{\circ}$
(labeled as ``Tel1'', ``Tel2'' and ``Tel3'', respectively),
facing the geometrical 
center of the arc of the exit window of the gas target.
The active area of each silicon detector was 50$\times$50~mm$^2$,
and the $\Delta$E layers were double-sided stripped silicon detectors
(DSSDs)
with 16 strips in two orthogonal dimensions with typical thicknesses of
30-60~$\mu$m, and the E layers were single-pad (SSDs) with typical 
thicknesses of 1.5~mm in which the protons were fully stopped.
The observed particles with the $\Delta$E-E telescopes 
are mostly identified as 
protons and alphas with sufficient resolution, 
together with a few deuterons and $^{3}$He ions. 
\begin{figure}[t!] 
 \centering
 \resizebox{8cm}{!}{\includegraphics{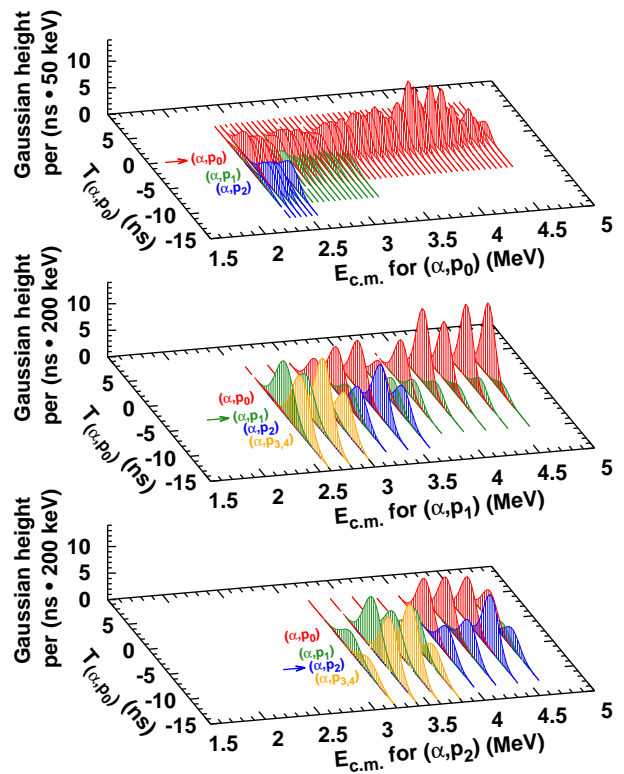}}
 \caption{\label{fig:gaus_fit}
 (Color online)
 Gaussian functions fitted to 
 the data of Tel1 with the higher-energy beam 
 at each calculated locus
 within each energy bin.
 The energy bin sizes are 50~keV for \apo\ and 200~keV for
 \api\ and \apii.
 The extracted excited state loci are
 indicated by arrows.
 %
 The same treatment has been also applied to the other
 telescopes, and the lower-energy data. 
 }
\end{figure}
%
\begin{figure}[t!] 
 \centering
 \resizebox{8cm}{!}{\includegraphics{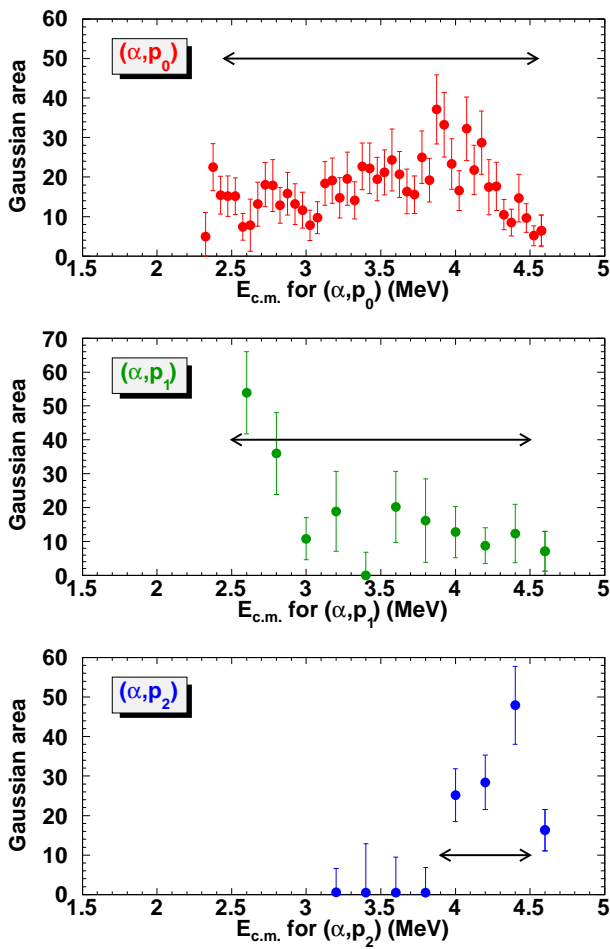}}
 \caption{\label{fig:counts}
 (Color online)
 Numbers of events of \apo, \api\ and \apii.
 Each value represents the area of the corresponding 
 Gaussian function extracted
 from those drawn in Fig.~\ref{fig:gaus_fit} 
 (Tel1, higher-energy run). 
 The arrows indicate the energy ranges selecting only the reliable data.
 %
 The same treatment has been also applied to the other
 telescopes, and the lower-energy data. 
 }
\end{figure}
%
%
\\ \indent
The present experimental setup 
is remarkable in terms of the
use of an extended gas target with a length of 140~mm along the beam axis
at a pressure of 400~Torr, which enables separation of
each transition to the $^{14}$N excited state
($E_{\rm x} = 2.313$~MeV, 3.948~MeV, etc.)
to be observed in different TOF 
between the first PPAC and the SSDs.
A similar idea was previously proposed 
in Ref.~\cite{PhysRevC.76.021603} 
and two 
experiments at CRIB \cite{Jung2012, Kim2015},
but none of them actually deduced cross sections
for excited-state transitions separately.
%
A more detailed explanation 
of this technique
was given by Ref.~\cite{Rogachev2010}
for separation of proton elastic and inelastic scatterings
with a $^7$Be beam.
However, again they did not demonstrate the TOF separation
of the actual data due to insufficient time difference
for the first excitation energy of $^7$Be 0.43~MeV, 
but showed the separation in the energy correlation between 
the light and the heavy recoil particles 
with the hybrid thick/thin target method.
Thus here we have for the first time made the full proof of principle 
for the \alphap\ reaction measurement with the TOF technique
in a comprehensive way.
%
%
%
\\ \indent
Figure~\ref{fig:ebeam_ep} illustrates how to determine 
the center-of-mass energies as well as the target position
for a reaction event,
under a given condition with 
a proton kinetic energy of 5~MeV,
a scattering angle of 40$^{\circ}$,
and the lower beam energy.
For the basis of this method, we directly measured 
the energy loss (red cross) and straggling (vertical error bar)
of the $^{11}$C beam particles
at several $^4$He gas target depths
with a silicon detector inside the gas,
and confirmed consistency with energy loss calculation 
\cite{Ziegler2010} shown as a decreasing function.
On the other hand,
the increasing functions represent possible kinematic conditions 
of different excited-state transitions for
a detected proton taking into account energy loss in the target gas
with the above given energy and angle.
The widths of these functions represent 
the systematic uncertainties.
Therefore one can determine the reaction energy and position 
simultaneously as the vertex of these functions.
%
The typical distance between 
reaction positions of \apo\ and \api\ calculated by this method
is about 70~mm, where the equivalent TOF difference 
becomes about 5~ns.
This TOF difference enables one to identify
which excited state in $^{14}$N the reaction reaches
event by event.
%
With a Monte Carlo simulation,
we estimated the accuracy
of the center-of-mass energy 45--65 keV depending on the
intrinsic energy resolution of each silicon telescope and 
the reaction angle; 
we found that 
the typical determination accuracy of reaction position is about 5~mm,
implying that the uncertainty in the solid angle is
eventually smaller than the statistical error we observed.
%
%
\\ \indent
Figure~\ref{fig:t-e_high-e_bg} shows the relation between
the TOF from the first PPAC to the SSD 
against the energy of protons detected in Tel1 
of the lower- and 
the higher-energy runs. 
The calculated TOF values are shown 
as solid lines for several allowed 
excited-state 
transitions,
of which the properties such as the onset of energy 
and the gradient 
nicely agree with the measured distributions in the plot.
%
The energy dependence of the time resolution of the SSDs 
was apparently independent from the choice of the beam condition,
thus it was estimated from the time broadening of 
\apo-only loci of the lower- and the higher-energy $^{11}$C beam data
and the $^{11}$B primary beam data.
%
The time broadening 
gradually increases from 1~ns to 3~ns as the proton energy decreases
from 12~MeV to 4~MeV,
and almost diverges very close to the energy threshold limited by
the $\Delta$E thickness, appearing in Fig.~\ref{fig:t-e_high-e_bg}.
Such a time broadening can be regarded as the intrinsic SSD resolution
as the effects of the proton
scattering angle or the beam energy broadening are expected
to be sufficiently small.
%
%
Although the separation of different excited-state loci 
was not perfect due to the time broadening,
we successfully extracted their mixing ratios
by fitting Gaussian functions to the time spectra within 
limited energy bins as described below. 
\\ \indent
Figure~\ref{fig:gaus_fit} and \ref{fig:counts} 
demonstrate the procedure to determine the number of events 
$N$ for each 
excited-state transition
from the same data (Tel1, higher-energy run)
as Fig.~\ref{fig:t-e_high-e_bg}.
%
We fitted those data 
projected to the time axis
by Gaussian functions
at each calculated peak within each energy bin.
The result of the fitting is shown in Fig.~\ref{fig:gaus_fit}.
Note that the origin of the time axis of Fig.~\ref{fig:gaus_fit}
is realigned to the calculated \apo\ locus,
which leads better separation between different loci
within a finite size of an energy bin.
The energy axes in Fig.~\ref{fig:gaus_fit} and \ref{fig:counts}
are converted to the center-of-mass energies $E\cm$
reconstructed with the $Q$-values for
\apo\ (top panel, $Q_0 = 2.923$~MeV), 
\api\ (middle panel, $Q_1 = 0.610$~MeV)
and \apii\ (bottom panel, $Q_2 = -1.030$~MeV), respectively.
The $E\cm$ bin sizes were 50~keV for \apo, and 200~keV for
\api\ and \apii.
%
Each excited-state locus in Fig.~\ref{fig:gaus_fit}
is labeled as \apo\, etc., 
and the arrows indicate which locus is extracted.
%
%
%
%
Figure~\ref{fig:counts} shows the 
extracted numbers of events of \apo, \api\ and \apii\
as the area of the corresponding Gaussian function
$N$ defined by
%
%
%
%
%
\begin{equation}
 N = \sqrt{2 \pi} \frac{H \sigma}{\Delta t_{\rm bin}}\ ,
\end{equation}
 where $H$ and $\sigma$ are the height and the width of the Gaussian
function respectively, and
$\Delta t_{\rm bin} = 1$~ns is the bin size of the normalized time
$T_{(\alpha,p0)}$.
%
Thus the error of the counts $\Delta N$, namely, that of 
the area of the Gaussian function can be defined as
\begin{equation} \label{eq:area}
 \Delta N = N \sqrt{ \left(\frac{\Delta H}{H} \right)^2 
  + \left(\frac{\Delta \sigma}{\sigma} \right)^2 }\ ,
\end{equation}
where $\Delta h$ and $\Delta \sigma$ are the error of the fitted parameters.
%
%
The arrows in Fig.~\ref{fig:counts}
indicate the energy ranges selecting only the reliable data,
eliminating the data at the ends of the energy ranges,
and the \apii\ data below the onset energy of $(\alpha,p_3)$
as they overlap considerably.
We repeated the same procedure for the data of the other telescopes,
and also for the lower-energy run.
%
%
%
\\ \indent
We also performed a measurement with the same setup but the 
gas target filled with argon gas of an equivalent thickness
in order to subtract background protons which reached only
the most forward telescope (Tel3) passing through the gas target windows 
from upstream in coincidence with the beam particles triggering. 
Such protons were also identified by the TOF-versus-energy 
information in a larger time range
and mostly separated from the \alphap\ reaction events. 
The only influential case was protons from the Mylar foil of the
second PPAC produced by elastic scattering of beam particles,
which partially overlapped with the \api\ and the \apii\
data of comparable intensity.
We only eliminated these data, otherwise 
the angular distribution of the background-subtracted data 
of Tel3 were basically consistent with the corresponding data 
of Tel1 and Tel2 within the errors,
which were taken into account to the angle-averaged cross sections
as discussed later.
\section{Astrophysical $S$-factors}
\begin{figure}[t]
 \centering
 \resizebox{8cm}{!}{
 \includegraphics{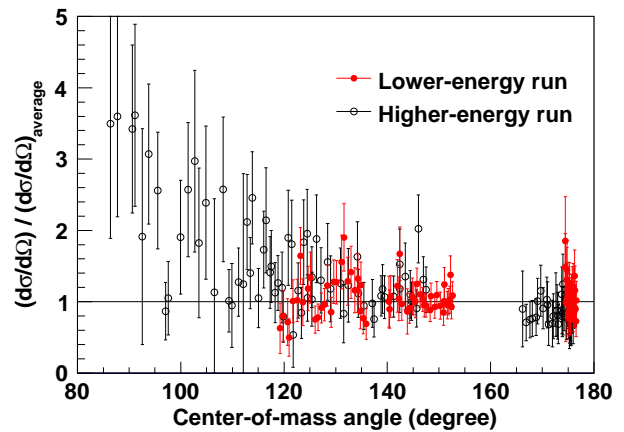}}
 \caption{\label{fig:ang_distr}
 (Color online)
 Ratio of the observed differential cross section 
 to that of the weighted average at each energy 
 versus the center-of-mass angle.
 The red solid circles are of the lower-energy run and 
 the black open circles are of the higher-energy run.
 }
\end{figure}

%
%
The present data appear to be nearly isotropic,
because 94\% of the data points are located within a factor of 2 or less
from the weighted mean value of the angular distribution spread 
at each energy.
%
This is shown up in Fig.~\ref{fig:ang_distr}
as the ratio of the observed differential cross section 
to that of the weighted average at each energy 
versus the center-of-mass angle. 
The angular distributions of the lower- and the higher-energy runs
in their overlapping range from 2.30~MeV to 2.75~MeV
were also in agreement within their errors, 
which confirms the self-consistency of our measurements.
%
\begin{figure*}[t!]
 \centering
 \resizebox{15cm}{!}{
 \includegraphics[angle=270]{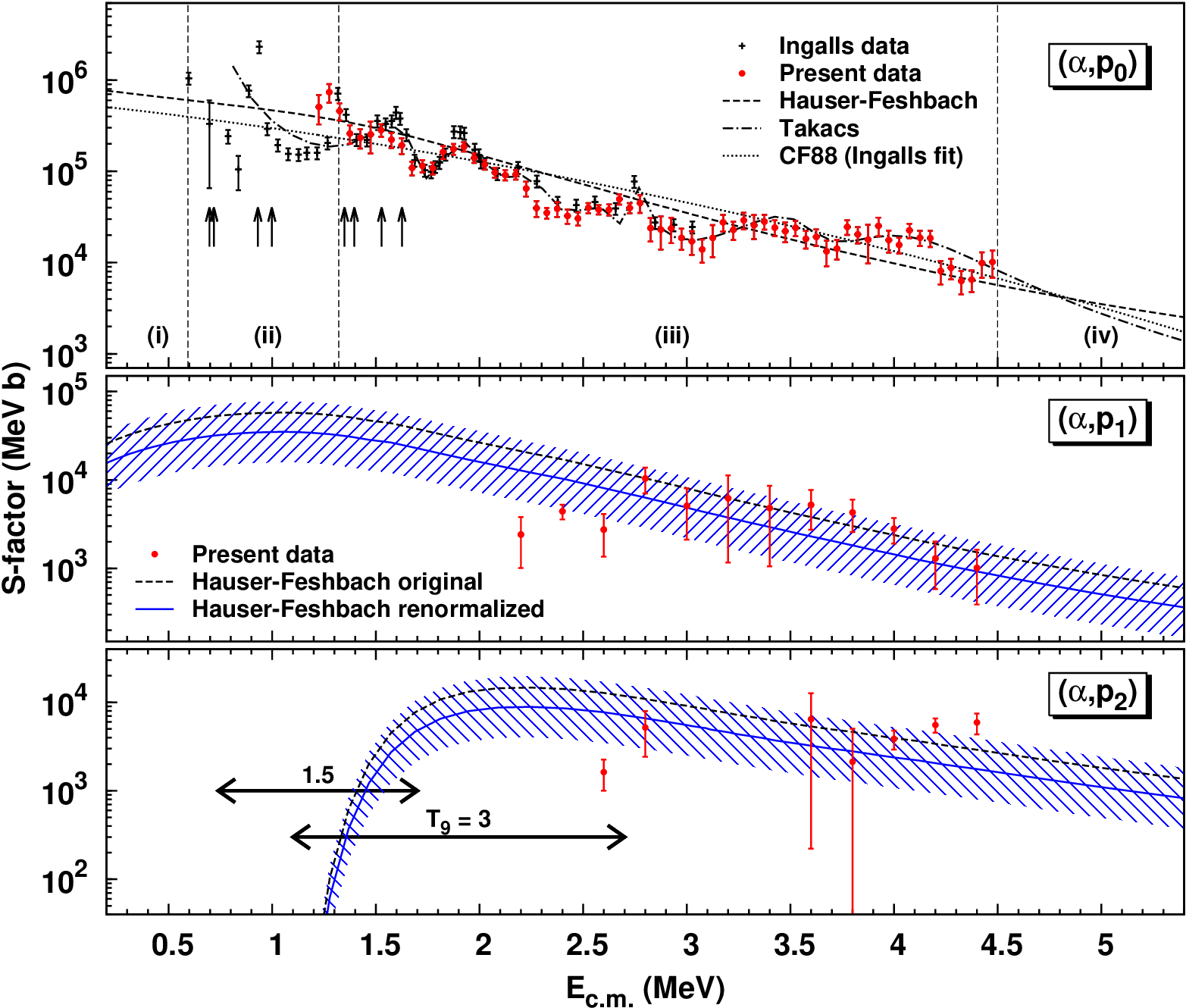}}
 \caption{\label{fig:s-factors}
 (Color online) $S$-factors for \capno\ (top), \capni\ (middle) 
 and \capnii\ (bottom).  
 The present data 
 of both the lower-energy run and the higher-energy run 
 are shown as red solid circles with error bars. 
 The original Hauser-Feshbach calculations for each excited-state 
 transition are shown as dashed curve, respectively.
 In the \apo\ panel, 
 the data of Ingalls {\it et al.}~\cite{PhysRevC.13.524} 
 (cross with error bars),
 and compilation curves of
 CF88~\cite{caughlan1988thermonuclear}
 and Tak\'acs {\it et al.}~\cite{Takacs2003169}
 are also shown for comparison,
 and
 the arrows indicates the most significant 8 energies
 of the known resonances below 1.7~MeV.
 %
 %
 For \api\ and \apii,
 the renormalized Hauser-Feshbach curves (solid curve)
 to the present data
 with their errors (hatch) are also shown.
 }
\end{figure*}
%
\\ \indent
In Fig.~\ref{fig:s-factors}, 
the newly obtained astrophysical $S$-factors 
of both the lower-energy run and the higher-energy run 
for \apo, 
\api\ and \apii\ are plotted as red solid circles with error bars.
%
The crosses with error bars are from the time-reversal reaction data 
\cite{PhysRevC.13.524} with the detailed balance theorem.
%
The Gamow windows of $T_9 = 1.5$ and 3 are indicated with arrows in the
bottom panel.
For comparison,
the Hauser-Feshbach model calculations for each transition
by the code \textsc{non-smoker}$^{\textsc{web}}$ \cite{Desc2006, raus_priv}
are shown as the dashed curve in the respective panels,
and the \apo\ $S$-factors of CF88 \cite{caughlan1988thermonuclear} 
(dotted curve) 
and that of Tak\'{a}cs {\it et~al.} \cite{Takacs2003169} (dashed-dotted curve)
are shown together in the top panel.
\\ \indent
The \apo\ $S$-factor by the Hauser-Feshbach calculation
roughly agrees with the present experimental data
within a factor of 2 over a several MeV range,
but obviously overestimates the rate at the lowest energies.
%
%
One can see that the present \apo\
data are mostly consistent with the previous data
by Ref.~\cite{PhysRevC.13.524} and \cite{Takacs2003169}
over most of the energy range 
except near the lower energy limit of the measurement
around 1.3~MeV.
%
%
The peak position in the present work near the known resonance at 
1.349~MeV looks shifted to a slightly lower energy.
%
At the lowest energies below $E\cm = 1.3$~MeV, 
the yield of the \apo\ events
rapidly dropped, and not all of the three telescopes
counted reliable \apo\ events.
Thus the angle-integrated cross section of these data
are lacking over some angular ranges,
and so we adopted the data only above that energy 
for the reaction rate calculation.
%
%
%
%
%
\\ \indent
The 8 arrows in the top panel indicate 
the positions of the known low-lying resonances 
with significant total widths,
corresponding well to the peaks appearing in the experimental data.
%
For known excited levels
\cite{ajzenberg1991energy}
in the corresponding energy region,
most resonance energies and total widths 
have been well determined via proton elastic and inelastic
scattering on $^{14}$N,
and some spins, parities and orbital angular momenta
were assigned by $R$-matrix analysis \cite{West1969,Radovic2008},
but no $\alpha$ partial widths were known.
%
Here we do not demonstrate $R$-matrix analysis on these data,
because we have only limited statistics for each resonances,
the resonance widths are comparable to the energy resolution, 
and the most resonances overlap each other,
which would unlikely lead any new information or
constraints beyond what was determined in the previous works.
%
We also observed the excitation function of the 
$\alpha$ elastic scattering down to $E\cm \sim 2$~MeV
with some resonant structure at the most forward angles
even below the Coulomb barrier of 3.5~MeV.
However, 
no useful information could be extracted 
in the lower energy range 
to which the reaction rate below $T_9 = 3$ is sensitive.
Thus we neglected to pursue further analysis of these data
in this paper.
%
%
\\ \indent
The present \api\ and the \apii\ cross sections come out
about one order of magnitude lower than the \apo\ one,
and those of the Hauser-Feshbach calculation
appear to be larger than the present experimental data.
For the reaction rate calculation, 
we extrapolated those $S$-factors toward the full energy range 
using the shape of the Hauser-Feshbach calculations
renormalizing to the present data 
by the logarithmic least-squares method.
%
Note that the utilization of the statistical model 
is not to justify its applicability
to such a light nuclear system, 
but rather to reveal difference of its absolute value. 
%
The renormalized Hauser-Feshbach curves 
are shown in addition to the original ones
in the middle and the bottom panels of Fig.~\ref{fig:s-factors}.
%
The normalization factors for \api\ and \apii\ 
are 0.61 and 0.69,
and their logarithmic standard deviations are 2.2 
and 2.6, respectively, as drawn with hatched areas.
%
\section{Reaction rates}
To derive the total reaction rate of \capn, 
we adopted 
new $S$-factors based on all presently available data 
as follows. 
The \apo\ data were selected 
from each energy section defined as (i)--(iv) by vertical dashed lines 
in Fig.~\ref{fig:s-factors};
(i)~the CF88 smooth curve, 
(ii)~the data points of the time-reversal reaction \cite{PhysRevC.13.524},
(iii)~the present data points,
(iv)~the compilation of Ref.~\cite{Takacs2003169} up to 25~MeV.
For \api\ and \apii,
we adopted the present data points for the measured energy regions, 
otherwise the renormalized Hauser-Feshbach
curves as extrapolation.
%
\begin{figure}[t]
 \centering
 \resizebox{8.cm}{!}{
 \includegraphics{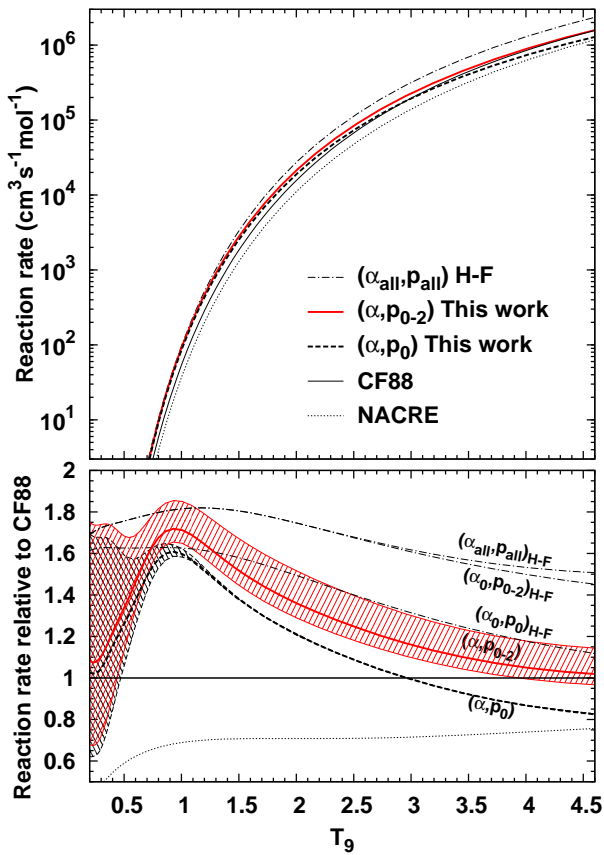}}
 \caption{\label{fig:rate_ratio}
 (Color online) 
 Absolute \capn\ reaction rates 
 of the present data and the currently available ones (upper) 
 and their ratios to the CF88 one (lower).
 %
 The uncertainties of the present \apo\ rate and
 total rate are drawn as hatches attached in the bottom panel.
 %
 }
\end{figure}
%
We performed numerical integration of those $S$-factors
based on Simpson's rule 
in energy ranges where the integrated values converged well.
%
For the integration,
the values at energies in between data points
were linearly interpolated from the neighboring values.
\\ \indent
Figure~\ref{fig:rate_ratio} shows 
the absolute reaction rates (upper)
and the reaction rates relative to the CF88 rate (lower)
of the present data and
several previously reported ones. 
%
%
 The hatched regions 
 in the lower panel indicate 
the errors of the present rates.
The error of \apo\ arises only from an uncertainty of
a factor of 1.65 in the energy section (i)
estimated as the logarithmic standard deviation of the present data
against the CF88 curve,
while those of \api\ and \apii\ from the uncertainties of the
renormalized Hauser-Feshbach functions.
In the $\nu p$-process temperature range ($T_9 = 1.5$--3)
\cite{WanajoJankaKubono2011},
the present \apo\ reaction rate
is enhanced from the CF88 rate
by about 
40\% 
at most, 
mainly due to the resonances around 0.9~MeV
and 1.35~MeV
which were not taken into account at all in neither CF88,
NACRE nor Ref.~\cite{Takacs2003169}. 
The contribution from the \api\ and \apii\ reaction rate
to the total reaction rate is about 20\% of the \apo\ at most,
and the error of the total reaction rate 
results in about 15\%.
The Hauser-Feshbach rate for \apo\ overestimates at low temperatures
and approaches to the experimental rates as the temperature increases,
as expected from the behavior of its $S$-factor in
Fig.~\ref{fig:s-factors}. 
The Hauser-Feshbach model calculation for the excited states shows an
insignificant difference between the \reac{}{}{\alpha}{p_{0-2}}{}{}\ rates
and the total rate \reac{}{}{\alpha_{\rm all}}{p_{\rm all}}{}{},
which suggests the present measurement up to \apii\ is sufficient.
\\ \indent
As for the astrophysical implications of the present work, 
it turns out that the enhancement of the \capn\
reaction rate by less than a factor of two
would not be a big impact to
change the relevant nuclear-burning scenarios;
%
by the present reaction rate,
the deviation of the branching condition 
between the \capn\ reaction and the $^{11}$C $\beta$-decay
from the previous condition \cite{wiescher1989hot}
is negligible,
as the neighbor branching conditions of other reactions
are located so far from that of this reaction
by many order of magnitude
in the density-energy dependency
investigated for metal-poor stars.
%
%
Therefore we conclude that the present result
could also support the validity of this reaction
as a possible path which breaks out from the $pp$-chain region 
to the CNO-cycle region in the $\nu p$-process \cite{WanajoJankaKubono2011}.
%
\section{Summary}
%
We have performed a direct measurement of 
the \capn\ reaction cross section at stellar energies of 
1.3--4.5~MeV 
by the extended thick-target method,
which enabled separation of both the \apo, \api\ and \apii\ transitions 
in time of flight. 
The present measurement provided one of the most comprehensive results
among relevant stellar \alphap\ reactions 
under explosive burning conditions.
%
%
The new total reaction rate lies
between the previous \apo\ rate
and the total Hauser-Feshbach rate, 
which still supports the validity of relevant 
explosive hydrogen-burning process scenarios 
such as the $\nu p$-process 
that it proceeds via the \capn\ reaction 
in addition to the triple-$\alpha$ process.
We expect that the present extended thick-target method
with the TOF technique 
is widely applicable to 
other $\alpha$- or even proton-induced reaction or scattering
measurements.
%
%
\begin{acknowledgments}
 We thank the accelerator staff at RIKEN and CNS,
 Prof. T.~Rauscher for his help with the \textsc{non-smoker} calculations,
 and Dr. S.~Wanajo for productive discussions
 on the $\nu p$-process.
 This work was partly supported by JSPS KAKENHI 
 (Grant No.~25800125 and 21340053).
\end{acknowledgments}

\end{document}